# Evaluating the Self-Optimization Process of the Adaptive Memory Management Architecture Self-aware Memory

Oliver Mattes
Institute of Computer Science & Engineering (ITEC)
Karlsruhe Institute of Technology (KIT)
mattes@kit.edu

Wolfgang Karl
Institute of Computer Science & Engineering (ITEC)
Karlsruhe Institute of Technology (KIT)
karl@kit.edu

*Abstract*—With the continuously increasing integration level, manycore processor systems are likely to be the coming system structure not only in HPC but also for desktop or mobile systems. Nowadays manycore processors like Tilera TILE, KALRAY MPPA or Intel SCC combine a rising number of cores in a tiled architecture and are mainly designed for high performance applications with focus on direct inter-core communication. The current architectures have limitations by central or sparse components like memory controllers, memory I/O or inflexible memory management.

In the future highly dynamic workloads with multiple concurrently running applications, changing I/O characteristics and a not predictable memory usage have to be utilized on these manycore systems. Consequently the memory management has to become more flexible and distributed in nature and adaptive mechanisms and system structures are needed. With Self-aware Memory (SaM), a decentralized, scalable and autonomous self-optimizing memory architecture is developed. This adaptive memory management can achieve higher flexibility and an easy usage of memory.

In this paper the concept of an ongoing decentralized self-optimization is introduced and the evaluation of its various parameters is presented. The results show that the overhead of the decentralized optimization process is amortized by the optimized runtime using the appropriate parameter settings.

## I. INTRODUCTION

After the former race for higher CPU frequencies, in recent years the performance improvements of microprocessors were achieved by combining an increasing number of CPU cores to yield into manycore processors. Advances in semiconductor technology made it possible to integrate multiple homogeneous cores in a tiled architecture, but also more and more systems with heterogeneous cores like GPUs or accelerators are available. But present manycore systems have restrictions by limiting components like a central memory controller, a quite limited number of external memory components or a static or semi-dynamic memory assignment. With that these systems are designed for some special application scenarios and restricted in usability and programmability, most of them for executing single high performance applications on several cores and using direct inter-core communication.

With the so called memory wall [1], the difference between the uprising CPU speed and the slow external memory, is getting more important with an increasing number of cores. So far, these systems commonly offer memory access over a small number of controllers to just one or a few external memory components, with limited I/O bandwidth and varying latencies from the different cores to the memory. We could confirm this behavior on a Tilera Tile-Gx platform with own measurements of access times and conflicts for different memory usage scenarios [2].

This lack in the memory system leads to inefficient memory assignment and causes congestion [3] getting worse scaling the core count or integrating heterogeneous cores. Future application scenarios consist of multiple high-dynamic and concurrently running applications. In most cases, an optimal initial assignment of memory to tasks is not feasible, caused by data locality issues, placement restrictions and memory regions, which are already occupied by other tasks.

In order to scale the memory with the rising core count and to tackle the problem of optimizing the management and assignment of memory to tasks in highly dynamic scenarios, we propose Self-aware Memory (SaM) [4]. In the following this scalable memory management system for adaptive computing systems is presented, which utilizes an ongoing self-optimization process without a central decisive instance, in order to get a continuous verification and optimization of the system behavior. The evaluation with a SystemC-based simulation, presents the results of iterations over the parameters of the optimization process. The results give advice for these parameters and their impact on the optimization and show that SaM and its adaptive memory management techniques can be realized with a limited management overhead, in return achieving higher flexibility and simple usage of memory in future system architectures.

The paper is organized as follows. After a short introduction to related work is given in Section II, in Section III an introduction into SaM is given. In Section IV the decentral self-optimization mechanism of SaM is presented, followed by the implementations and evaluation scenario in Section V and VI, the evaluation of the parameters of the decentral monitoring and their impact on the optimization in Section VII and the conclusion in Section VIII.

## II. RELATED WORK

In recent years the number of cores per processor was increased more and more. In research and industry some first manycore systems came up, but so far only a few of them are commercially available. Examples for these first Manycore systems are the KALRAY MPPA (Multi-purpose Processor Array) [5], the FPGA-based RAMP Blue [6], and the Intel SCC (Single-chip Cloud Computer) [7] as well as the Xeon Phi coprocessor cards, which are based on the Intel Larrabee [8]. The commercially available Tilera TILE-Gx [9] manycore processors are a follow-up of the MIT RAW project [10], in which the basic principles of a tiled architecture were developed. Most of the first manycore systems are build in this tiled architecture, in which multiple smaller, mostly homogeneous cores are connected over networks on chip (NoCs) and combined on a single chip. But due to restrictions in their strong centralistic design, most of these systems are limited to execute parallel applications like streaming applications, which mainly communicate directly between the cores, therefore using small on-chip memories. Access to the external memory is achieved over one or only a few external memory controllers with limited I/O bandwidth and varying latencies from the different cores to the memory. The difference between the uprising CPU speed and the slow external memory is getting more important with an increasing number of cores.

We could confirm this behavior on a system with a Tilera Tile-Gx 8036 processor with own measurements of access times and conflicts for different memory usage scenarios [2]. In this system with 36 cores, access to the two external memory modules is achieved via the grid NoC over two I/O links each. We measured the access latencies from each core to a single memory module using it as private or shared memory or for message passing. The measured memory access times strongly depend on the position of the tile in the grid and its assigned and accessed memory modules. As expected for currently available manycore systems, the access times are at a minimum 13,5 % higher for the farthest compared to the nearest tile. Moreover, the value for this slowdown can only be achieved when the whole system and network can be exclusively used by a single core. In real application scenarios, this actually will never be the case. With rising core count the slowdown instead will rise to a much higher level due to mutual interference using the same network or in accessing the same memory component.

Autonomic or organic computing with integrated self-x functionalities is a research area, which was tackled since the last decade. A visionary overview was given in [11] in which the structure of autonomic elements was described as basic principle of self-managing systems. An autonomic manager and a managed component build up an autonomic element. In the manager the so called MAPE cycle is executed to monitor, analyze, plan and execute the management task, based on information by the underlying self-knowledge. This principle is captured by the decentral memory system and self-optimization process of SaM.

In [12] associative counter arrays are introduced to accumulate and preprocess monitoring information. In case of an overflow of a counter a status message is sent to the next upper monitoring instance in the hierarchy, up to a central instance which processes the collected information and initiates a reaction. Instead of a centralized management instance, in the presented work, the monitoring and the subsequent optimization process is handled by a cooperation of decentral self-managing components. Therefore new decentral ways for discovering and calculating optimization possibilities as well as a consensus building process have to be figured out.

### III. SELF-AWARE MEMORY

Self-aware Memory (SaM) [4] is a decentral memory architecture, in which the memory is split up into several self-managing components. The initial intention was to build up a memory architecture without a central management instance as a single point of failure and for scalability reasons. Now, the main goal of SaM is to develop an autonomous memory subsystem for increasing the overall system reliability, flexibility and adaptability, which is crucial for upcoming computer architectures and the proposed high-dynamic workloads.

Within SaM, the memory is split up into independent units. No initial assignment to a specific core is needed. As the self-awareness in the name implies, these components each collect information about their state, e.g. allocation, load, or condition, which then is used for the self-management. In total, the SaM components interact as a distributed and extended memory management unit and control memory allocation, access rights, and ownership in a distributed manner.

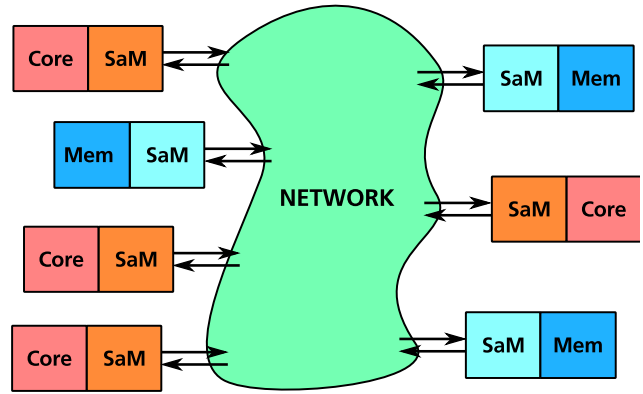

Fig. 1.  Distributed SaM structure with assigned management components

SaM is built as a service-oriented architecture, in which the memory modules offer their service of data handling to processor cores. The structure of SaM is shown in Figure 1. The memory is divided into several autonomous self-managing memory modules, each consisting of a SaM management component and a part of the physical memory. Each management component handles access and mapping of its attached physical memory, and administrates free and reserved space. As a counterpart, a SaM management component is assigned to each compute core to augment the core with self-management functionality, acting as an enriched MMU. It is responsible for handling memory requests, performing access rights checks, and mapping of the virtual address space of the connected core into the distributed SaM memory space. The virtual to physical address translation as well as the whole memory allocation and management is realized in a transparently for the compute cores.

SaM enables access to private as well as shared memory and integrates efficient synchronization techniques [13]. The memory system also can guarantee the Transactional Memory (TM) principles atomicity, consistency and isolation in a combined HW and SW approach. This provides an easy and flexible way to program access to shared memory and an abstract view of the memory resource.

As a side product, SaM enables thread creation and management as well as allocation of compute nodes without a central management instance [14] with a POSIX-like thread model.

In total the strict decentralization enables a highly scalable and fault tolerant memory system which can keep up with the increasing number of compute cores. Beyond that the interaction of the SaM management is hidden, so the compute cores have access to an abstract memory resource without restrictions in programmability and usage. To enable adaptivity to the proposed high-dynamic application scenarios, SaM includes a decentralized self-optimization mechanism, which is explained in more detail in Section IV.

## IV. Decentralized Self-optimization

The proposed high-dynamic application scenarios with multiple concurrently running applications, changing I/O characteristics and a not predictable memory usage, call for mechanisms to adapt to changing needs. The distributed structure of SaM fits well to this dynamic scenario, because different applications can be handled by independent parts of the self-managing components. To enable adaptivity, a decentralized self-optimization mechanism was integrated [15], [16] in SaM.

Autonomic computing, more precisely the working steps of autonomic elements, are grounded on the MAPE cycle [11]. With the scenario of self-optimization in a decentralized system, the four steps of the MAPE cycle could be mapped on corresponding sub working steps. Concerning the decentralization an additional step, the consensus building, is included to handle the decentralized agreement on optimization proposition between the involved components. Basis of all steps is the knowledge. Here it means, that each component knows its own status – it is self-aware – and for example provides information about its usage, allocation of the memory regions, health situation etc.

### A. Self-Optimization Cycle

Our proposed optimization cycle is based on the following five steps:

1) Decentralized Monitoring and Data Preprocessing: local data collection per system component and periodic exchange with neighbors.
2) Data Analysis: analysis of the monitored information, including associative counters, which provide a threshold value for the following optimization step.
3) Optimization Algorithms: initiated by the overflow of an associative counter, in this step an optimization proposition is calculated using a dedicated optimization algorithm.
4) Decentralized Consensus Building: Validation of the optimization proposition and decentralized voting procedure.
5) Optimization: The actual execution of the accepted proposition. Depending on the optimization algorithm this might be a data migration process combined with an update of the address management tables. The virtual memory addresses on CPU side are not modified.

After these steps an optimized system behavior is achieved for the moment. This is an coincident optimization process, ongoing on all system components to react on dynamic changes.

In this paper we focus on the different parameters of the first step and their influence on the result of the optimization process. In [15], [16] more details to the other steps can be found.

### B. Global vs. Local Optimization

The presented optimization process is aligned to several concurrent local optimizations of the distributed self-managing components. A global optimization could be reached but is not the main target, because due to the high-dynamic application scenario, a stable global system state is not available. An continuous and concurrent process of multiple decentral local optimizations here leads to a higher flexibility and reliability.

This approach is theoretically justified by the decentralized decision making for multi-agent systems [17], describing decision making with several instances, called agents, and negotiations without any central instance. As explained in [15], [16] for optimizations the system state is saved in decision vectors. Concerning the ongoing refinement of the multiple vectors, the decision information from the system components could be outdated, which is why it is often not possible to find a global optimal decision.

Regarding the highly dynamic workload this problem is enhanced, because all system changes result in updated decision vectors and a stable global decision vector is not achieved. A system-wide information distribution also does not scale with rising system size, because the number of necessary messages is getting too big. Therefore a global optimization is not reasonable for the assignment. In the underlying scenario, several applications are also locally bound to a distinct part of the system, in which then a local optimization could be done.

### C. Optimization algorithms

With the presented optimization process several optimization algorithms can be deployed. To adapt a distinct algorithm, parameters like the exchanged monitoring info in step 1, the threshold of the associative counters in step 2 and the executed optimization in the last step of the process have to be adjusted. Additional parameters, e.g. the migration costs to weigh up the optimization advantages with the costs for the optimization, are used by the optimization algorithms.

Several algorithms are conceivable to define the optimization. To take as examples we address some optimization algorithms in here: Designated target of the latency optimization is to reduce the distance between the compute node and the actually used memory region by relocating the memory to a better located memory component. A reduced distance depends on the network structure, lower latencies or higher bandwidth of different connections. Concerning load balancing, a memory component observes a high load and tries to scatter memory regions to other memory components in order to prevent an upcoming congestion. As an example for reliability optimizations the self-awareness could be used to count ECC errors at data accesses in a memory component. If a threshold is exceeded the memory component tries to save its data by spreading it to other memory components. As last example energy saving is of high importance. Temporarily powering down designated system components could be achieved if the data of memory components is evacuated and agglomerated in less components. Within this scenario, the amount of energy which is additionally used by the SaM components has to be compared with the saved energy by the optimized system.

### D. Decentralized Monitoring

Basis of the self-optimization process is the knowledge of the current system state. As the first step in the optimization process, the independent system components have to collect information about themselves and their neighborhood, e.g. the status changes and its memory usage.

First of all, each component has to collect information about itself. In an ongoing exchange with their neighbors, the stand-alone components cumulate information. Integrating these additional information in their own status messages, the information and the view of the system grows step-by-step on each distributed component.

In periodical intervals information about the condition of the components is then exchanged between neighbors. Depending on the available network type this could be done in different ways. In the following, explicit messages are used to distribute the information to a group of surrounding neighbors within a specific number of hops. Multiple hops lead to a wider system knowledge for more global optimizations, but with a higher number of monitoring messages. Less hops reduce the number of monitoring messages on the network, but the optimization only could be done on a bounded local region. A trade-off between these two principles has to be done. In figure 2(a) two examples of neighborhoods with a radius of one (red) and two (blue) are provided.

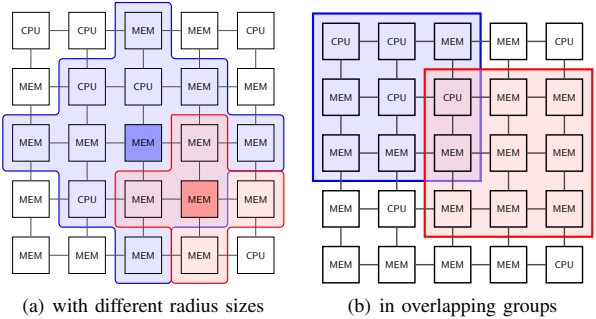

(a) with different radius sizes    (b) in overlapping groups

Fig. 2. Distribution of information

The emission period of these status updates can be varied. With a shorter period system changes are propagated faster, but this also leads to a higher amount of monitoring messages.

Associative counter arrays are used to arrange and pre-validate the collected information. If a threshold of an associative counter array is exceeded, the optimization algorithm is called, which then calculates an optimization proposition.

To treat only commonly used events, the monitoring cycle periodically sets back the associative counters.

## V. IMPLEMENTATIONS AND PROTOTYPES

Up to now there are 4 different evaluation prototypes for Self-aware Memory. The most common and flexible prototype is a SystemC-based simulation [4], [15], [16], which easily can be parameterized and adapted to several test scenarios and system structures. With this, the memory management mechanism and the self-optimization process was evaluated and developed. In addition a coarse-grained implementation using several FPGA boards [14], [13], each representing a CPU or memory component, connected over Ethernet is available. The third prototype exists as a SW daemon, running on normal PCs and redirecting memory access. It also can be connected to the FPGA-based version via Ethernet. In the context of the latency measurements we implemented SaM as a SW layer on the Tilera platform [2] to exemplarily demonstrate a high-dynamic and adaptive memory management on a existing manycore system. The following evaluation was done using the flexible simulation environment.

## VI. EVALUATION SCENARIOS

The evaluations in the following section are done using the SystemC-based simulation. As explained in the introduction, the usage scenario of upcoming manycore systems lies in several dynamically approaching concurrent applications, each running on a part of the system. For the evaluation various possible system configurations have been examined in permuting the parameters of the optimization process in repeated simulations.

In order to reproducibly evaluate the optimization process, initial usage scenarios and system states have to be provided, which could be replayed. As initial usage scenario, several different randomly chosen tasks are scheduled and distributed over the system. To start point of the actual evaluation time frame can be seen as a snapshot of a system running such a dynamic workload. Up next a new application is scheduled and the allocation of memory regions results in unsuitable located parts of the system. The purpose is to identify the unsuitable state and to initiated an optimization by the self-managing system components. In the following evaluations as the optimization algorithm a locality optimization was executed to migrate the memory regions closer to the cores on which the task is executed.

This work is motivated to enable a flexible memory management for high-dynamic application scenarios. Up to now there are no predefined benchmark scenarios for manycore systems available. To simulate these dynamic application scenario, we use a collection of memory traces we got from several benchmarks. This allows us to replay exactly the same scenario several times with changed parameters. Along with that it enables us, to easily run different evaluations with variable program and memory access phases. For each tile representing a CPU core, an application scenario using a sequence of traces is provided. The traces then are randomly chosen, representing an external triggered task, which is executed on a particular core. Different application scenarios can be simulated using schedules with a mixtures of these traces.

Same test scenarios are also executed with switched off self-optimization to be able to measure the influence and improvement. As a result the overhead of the optimization process can be compared with the accelerated program run-time.

## VII. EVALUATION RESULTS

In this section the results of multiple evaluation runs, permuting over the different parameters of the self-optimization process, are presented and rated concerning their impact on the optimization goal (here a locality optimization was executed to migrate memory pages).

### A. Monitoring cycle period

To treat only commonly used events, the monitoring cycle periodically sets back the associative counters and the optimization starts anew. The following evaluation results were obtained using a monitoring cycle period of 5000 simulation cycles. We also evaluated the same scenarios and parameter settings with monitoring cycle periods of 1000 and 10000. Their results closely resemble the here presented ones, so the optimal parameter settings are only changed in the exact values. Depending on the period, the values vary, at which the threshold and the emission period reach their limits.

### B. Threshold value

Associative counter arrays are used to arrange and pre-validate the collected information. If a threshold of an associative counter array is exceeded, the optimization algorithm is called, which then calculates an optimization proposition. As can be seen in Figure 3 and 4 the number of executed

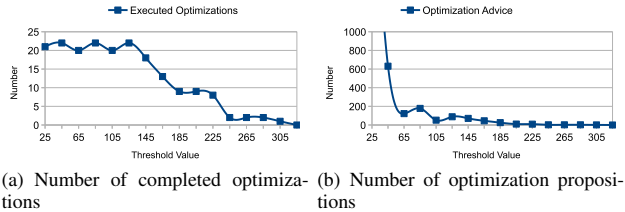

(a) Number of completed optimizations  (b) Number of optimization propositions

Fig. 3. Impact of the threshold on the number of optimizations

optimizations decreased with rising threshold values. As well,

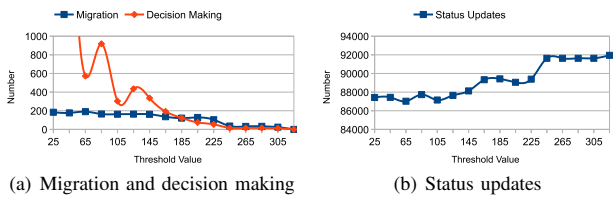

(a) Migration and decision making  (b) Status updates

Fig. 4. Impact of the threshold on the number of messages of the optimization process

for very small threshold values, the optimization algorithm is triggered too often, resulting in a very large number of optimization propositions and a concomitant high number of messages in the decision making process. Depending on the optimization algorithm, a higher oscillation of the number of these messages can be seen for middle threshold values.

### C. Emission period

The emission period of the status updates between the decentral components can be varied. With a shorter period system changes are propagated faster, but this also leads to a higher amount of monitoring messages. In Figure 5 and 6 results of the same evaluation setup are shown, using a fixed threshold of 45 but varying the emission period of monitored status updates. The number of executed optimizations is equal with changing emission periods. But the number of optimization propositions is decreased with longer periods due to less status update messages. Here, the time between emissions should be higher to save messages for decision making and status updates. As for varying the threshold value, the number of messages in the decision making process oscillate strongly for mid-values of the emission period without impact on the number of executed optimizations.

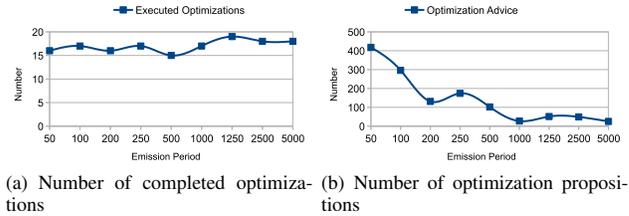

(a) Number of completed optimizations  (b) Number of optimization propositions

Fig. 5. Impact of the emission period on the number of optimizations

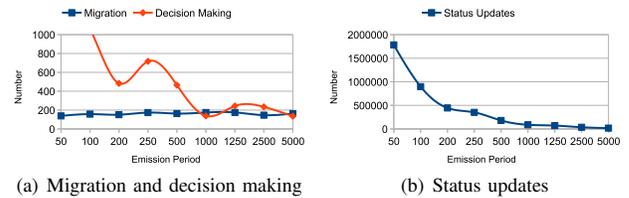

(a) Migration and decision making  (b) Status updates

Fig. 6. Variation of the message emission period

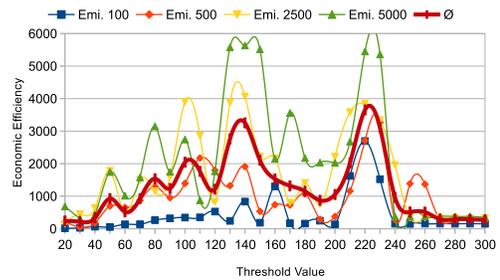

Fig. 7. Economic efficiency of parameter configurations of SaM

### D. Overall relations - Economic Efficiency

As can be seen in these examples, changes in different parameters can have opposing impacts on the optimization process. With the economic efficiency points out when the overhead of the optimization process is amortized and the optimized outweighs the unoptimized runtime. The lifetime of the broadcast for status updates, the neighborhood, is set to 1 or 2. Figure 7 shows the economic efficiency with a neighborhood of 1. No single best overall parameter configuration could be provided for that, but the optimization process runs faster and fine-grainer for the maxima with smaller thresholds. With a neighborhood of 2 the best optimization results can be achieved on average for medium threshold values.

## VIII. CONCLUSION AND OUTLOOK

In this contribution a decentralized autonomous memory architecture with self-optimization capabilities was presented. The introduced mechanisms enable the continuous verification and optimization of the management and assignment of memory to already running applications in a system without a central decisive instance. Depending on the collected information basis, several concurrent local optimizations could be performed, resulting in a better performance of the system. Regarding the assumed highly dynamic application scenario only parts of the system are used for one particular application. Along with the rapidly outdated decision information and the not scalable amount of monitoring information in a central instance, multiple local optimizations are favorable to global ones. Potential and temporary disadvantages for individual applications are going to be detected and re-optimized by the persistent optimization process.

The evaluation of multiple possible parameter configurations using traced benchmarks within a SystemC-based simulation showed, that a decentral memory management is feasible an the optimized runtime outweighs the small and manageable number of additional messages. Depending on the used optimization algorithm, the method during the decision making process, the neighborhood used for monitoring (global vs. local optimization), the threshold of the associative counters and the monitoring periods, choosing a suitable set of parameters limits the increase of traffic on the available network (NoC). By usage of systems with hybrid photo-electronic networks, it is also conceivable to split the fine-grained data transfers (e.g. monitoring and optimization messages) on electronic from the coarse-grained data transfers on photonic networks.

This work presents the evaluation results of latency optimizations. Nevertheless, the presented autonomous optimization mechanism could be reused for several other optimization purposes by integrating different optimization algorithms and metrics, e.g. load balancing, fragmentation or energy consumption. In this sense also the reliability of the system could be improved by the coordinated interaction of the self-managed system components.

To further improve the optimization results, aspects from machine learning in combination with program phases are being examined. Different caching mechanisms like CloudCache [18] have to be compared or combined with SaM and its own caching, used within the previously addressed transaction based synchronization mechanisms. An combination or integration for mutual enhancement of caching and external memory assignment could also be possible. Adapting and evaluating the mechanism for new and upcoming memory connections like 3D-stacked memory, photonic or hybrid photo-electronic networks associated with the changing system structure is another challenging and interesting step on our agenda.